\documentclass[aps,prl,floats,floatfix,twocolumn]{revtex4}

\usepackage{ifthen}
\usepackage{ifpdf}
\usepackage{color}

\ifpdf
\usepackage{graphicx}
\fi

\usepackage{latexsym}
\usepackage{amsmath}
\usepackage{amsfonts}
\usepackage{amssymb}
\begin{document}

\def\applss{\,\lower0.5ex\hbox{$\sim$}\kern-0.79em\raise0.5ex\hbox{$<$}\,}
\def\appgtr{\,\lower0.5ex\hbox{$\sim$}\kern-0.79em\raise0.5ex\hbox{$>$}\,}

\title[Chaotic spin pump]{Classical and quantum chaotic angular-momentum pumps}

\author{T.~Dittrich$^{1}$ and F.~L.~Dubeibe$^{2}$}

\address{$^{1}$Depto.\ de F\'\i sica, Universidad Nacional de Colombia,
and \\ CeiBA -- Complejidad, Bogot\'a D.C., Colombia\\ $^{2}$Facultad de Ciencias Humanas
y de la Educaci\'on,
Universidad de los Llanos, Villavicencio, Colombia}
\begin{abstract}

We study directed transport of charge and intrinsic angular momentum by periodically driven
scattering in the regime of fast and strong driving. A spin-orbit coupling through a kicked magnetic field confined to a compact region in space leads to irregular scattering and triggers spin flips in a spatially asymmetric manner which allows to generate polarized currents. The dynamical mechanisms responsible for the spin separation carry over to the quantum level and give rise to spin pumping. Our theory, based on the Floquet formalism, is confirmed by numerical
solutions of the time-dependent inhomogeneous Schr\"{o}dinger equation with a continuous source term.

\end{abstract}

\maketitle
\emph{Introduction.--}  The possibility to generate directed transport is one of the surprising applications of nonlinear dynamics, in dissipative as well as in Hamiltonian systems \cite{Tho83,AG99,Rei02}. In the absence of dissipation, it arises if phase space is strongly inhomogeneous, such as, e.g., in a mixed dynamics, and all binary spatio-temporal symmetries are broken that would lead to the occurrence of counter-directed trajectory pairs \cite{FYZ00}. The mechanisms of nonlinear transport generated  by mixed or chaotic Hamiltonian dynamics have been elucidated in systems with an extended periodic potential as in crystalline solids, dubbed ``ratchets''  \cite{SO&01,SDK05}, and in ``pumps'' \cite{DGS03,DD08,CDS12}, conceived as periodically driven scatterers inserted between two asymptotes of free motion. Hamiltonian systems are amenable to direct quantization using the Floquet formalism, possibly combined with Bloch theory. The resulting quantum ratchets  \cite{SO&01,SDK05} and pumps \cite{CDS12} exhibit directed transport owing to similar dynamical mechanisms as in the corresponding classical systems or even exploit genuine quantum effects without classical counterpart.

New phenomena emerge when internal freedoms are included, beyond the 
extended spatial coordinate where transport occurs. They enrich the dynamical scenario, in particular they can render integrable systems chaotic. At the same time, inner freedoms can take part, e.g., as ``passive scalars'', in  directed currents. An important application is generating polarized currents, an indispensable resource in spintronics \cite{Pri98}. Inducing directed spin transport by means of chaotic pumps is an attractive option to be addressed in this paper.

Polarized currents have been studied in the framework of ratchets in a variety of settings modelling extended solid-state or molecular systems, exploiting their static structural and electronic features \cite{SP&07,SB&10}. By contrast, we here consider transport of angular momentum owing exclusively to nonlinear dynamics. As a complementary orientation besides spin ratchets, we take conditions and basic features of the chaotic pumping of point particles into account \cite{DGS03,DD08,CDS12}: Working in the non-adiabatic regime of fast and/or strong external forcing permits, for example, to surpass linear response and to obtain directed transport already with a single driven parameter \cite{CDS12}.

We shall introduce angular-momentum pumps on the classical level, partially reviewing material from \cite{DD08}, to specify models, fix notations, and discuss typical dynamical scenarios: Kicked (impulsively modulated) magnetic fields, constrained to compact regions in space, provide the necessary spin-orbit coupling and intrinsically break time-reversal invariance (TRI), yet are simple enough to facilitate analytical and numerical treatments. We skip the semiclassical regime, as concerns the inner freedom, of large quantum angular momenta (in the following we use the terms ``angular momentum'' and ``spin'' interchangeably wherever no confusion is caused) and jump directly to the opposite limit of directed transport of spin-$\frac{1}{2}$-particles in quantum pumps, by driven chaotic scattering of charge carriers. We use Floquet theory to quantize periodically driven scattering without the limitations of adiabatic or perturbative approaches \cite{How79,Yaj79}, and present numerical evidence for directed spin transport, obtained by solving the time-dependent Schr\"odinger equation with source term \cite{JC&08,Dub10}. At least as concerns the underlying dynamics of the external degree of freedom, similar mechanisms apply as in the classical case. Pertinent features of chaotic pumps, such as the sensitive parameter dependence and frequent sign changes of the current, carry over to the quantum level and enable, in particular, pure spin without charge transport and v.v. We conclude pointing out the possibility of realizing chaotic spin pumps in the laboratory using present-day technology to drive quantum dots in the THz regime \cite{GK&93}. A particular topic left open by our work is the semiclassical analysis of the electron spins coupled to an orbital motion close to the classical limit \cite{LF91,CS&04,Cul09}. 

\emph{Classical angular-momentum pump.--} In order to model chaotic scattering of point particles with angular momentum, we seek drivings that simultaneously fulfill three tasks, (i) coupling the angular momentum to the orbital freedom such that the interaction goes beyond mere precession and not even the polar angle of the spin is conserved, (ii) rendering the dynamics at least partially irregular, and (iii) break TRI. Moreover, they should be sufficiently simple to facilitate their analytical treatment and permit efficient numerical simulations where necessary. These conditions are satisfied by magnetic fields modulated by a chain of delta kicks allowing for a reduction to discrete time, and confined to compact regions in the extended coordinate. In charged particles, the intrinsic angular momentum gives rise to a magnetic dipole moment, coupled to the spatial motion via the inhomogeneous magnetic field. Transversal components of the Lorentz force are neglected, assuming the longitudinal velocity to be sufficiently small. The force exerted on electrons by the induced electric field, for the model specified below, proves to contribute a term identical to the Lorentz force and is neglected for the same reason.


We consider one-dimensional spatial motion with momentum $p$ in the longitudinal direction $x$ coupled to a spin vector $\textbf{s} = (s_x,s_y,s_z)$ with components also in the  $(y,z)$-plane, by the Hamiltonian \cite{DD08}
\begin{align}
&H(p,x,\textbf{s};t) = H_0(p) + V(\textbf{s},x) \!\!\!\!\! \sum_{n=-\infty}^{\infty}\delta(t-nT-t_{\rm{in}}),\label{clhamiltonian}\\
&H_0(p) = \frac{p^{2}}{2 m_0},\,\,\,V(\textbf{s},x) = \gamma\,\textbf{s}\cdot\textbf{B}(x) \label{kinpot}
\end{align}
with mass $m_0$ and gyromagnetic ratio $\gamma$. Alternatively, the Lorentz force could couple the spin to an electric field, as in a Rashba term $\sim (\textbf{s} \times \textbf{p}_x)\cdot \textbf{e}_z$ \cite{SB&10}. The magnetic field is modulated in time by periodic kicks with period $T$ and phase $\phi_{\rm{in}} = 2\pi t_{\rm{in}}/T$ \cite{DGS03} and in space by an envelope $f(x)$ that vanishes outside the interval $[-a/2,a/2]$ yet is infinitely often differentiable within,
\begin{align} \label{bellies}
&\textbf{B}(x) = (0,B_1(x),B_2(x)), \,
B_{\sigma}(x) = A_{\sigma} f\left(x - (-)^{\sigma} \frac{a}{2}\right), \nonumber\\
&\sigma = 1,2,\,\,\,f(x) = \exp\left[\frac{-1}{(a/2)^2 - x^2}\right] \Theta\left[a/2-|x|\right].
\end{align}

\begin{figure}[h!]
\begin{center}
\includegraphics[width=7cm]{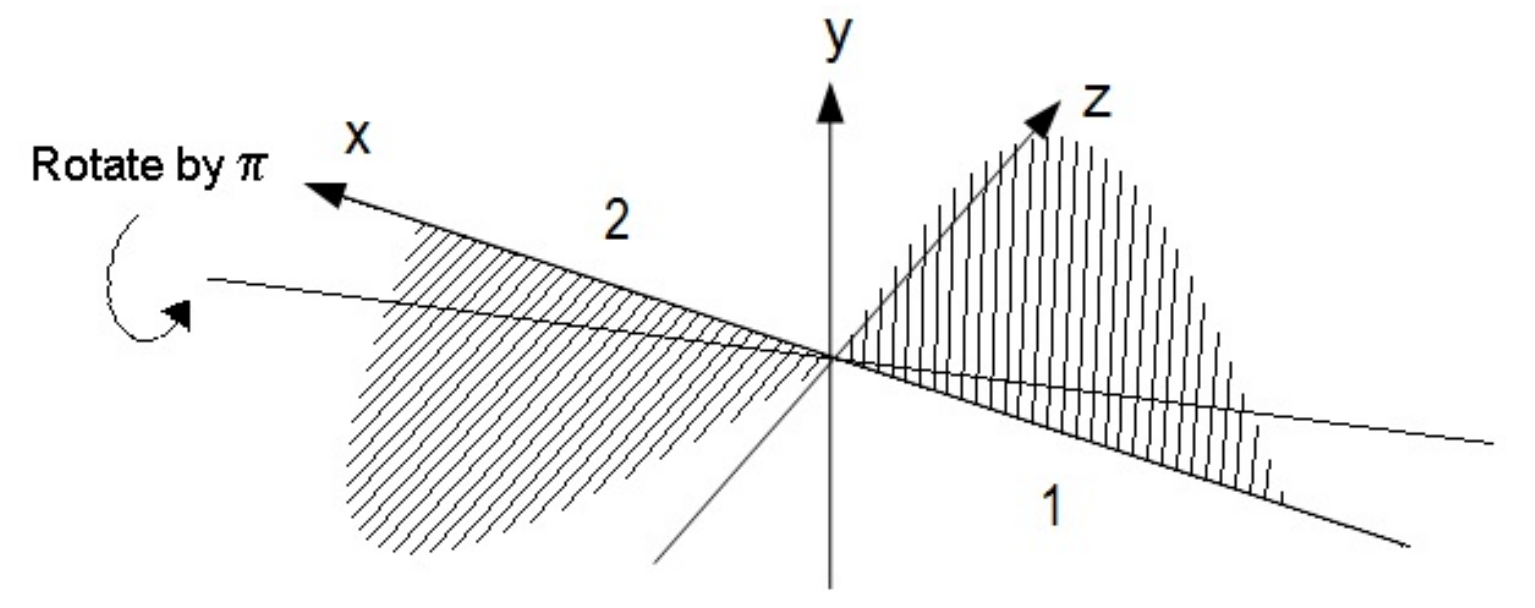}
\caption{Configuration of the magnetic field,
Eq.~(\protect\ref{bellies}). In each of the two
sectors $\sigma = 1,2$, the field is isotropic, with an angle of $\pi/2$
between them. For identical envelopes, the field is symmetric
with respect to rotation by $\pi$ around the bold line, corresponding to the transformation
$x \to -x$, $y \leftrightarrow z$.}
\label{Bconfig}
\end{center}
\end{figure}

Henceforth  we set $m_0 = \gamma = T = |\textbf{s}| = 1$. Where not specified otherwise, $a = 4$. We place time sections immediately before each kick, $t_n = nT - 0^+$, and align the $y$-axis as well as the axis of reference for the angular momentum, $\textbf{s} = |\textbf{s}| (\sin\theta\sin\varphi,\cos\theta,\sin\theta\cos\varphi)$, with the local field in each sector, $\textbf{e}_y \equiv \textbf{B}/|\textbf{B}|$, to arrive at stroboscopic maps for the two field sectors
\begin{align}
 p_{n+1} &= p_{n }- 2 x_{\sigma,n} B_\sigma(x_{\sigma,n}) \frac{\cos\theta_{\sigma,n}}
 {[(a/2)^2 -x_{\sigma,n}^2]^{2}}, \label{mapa}\\
 \varphi_{\sigma,n+1}&= \varphi_{\sigma,n}- B_\sigma(x_{\sigma,n})\,, \label{mapb}\\
 x_{\sigma,n+1}&= x_{\sigma,n}+ p_{n+1}\,, \label{mapc}
\end{align}
Here, $B_\sigma(x)$, $\varphi_{\sigma,n}$, and $x_{\sigma,n}=x_{n}-(-1)^\sigma a/2$ refer, respectively, to the magnetic field, azimuth, and position in each sector of the
interaction region. Precession within each field sector conserves the polar angle $\theta_n$, but passing from one to the other \cite{DD08}, it may change.

The spin-orbit coupling already violates TRI. In order to break also a remaining spatial symmetry under the rotation $x \to -x$, $y \leftrightarrow z$
(cf.\ Fig.~\ref{Bconfig}), we allow for a difference $\Delta A = A_2 -A_1$ as symmetry-breaking parameter. 

\begin{figure}[h!]
\begin{center}
\includegraphics[width=5cm, angle=270]{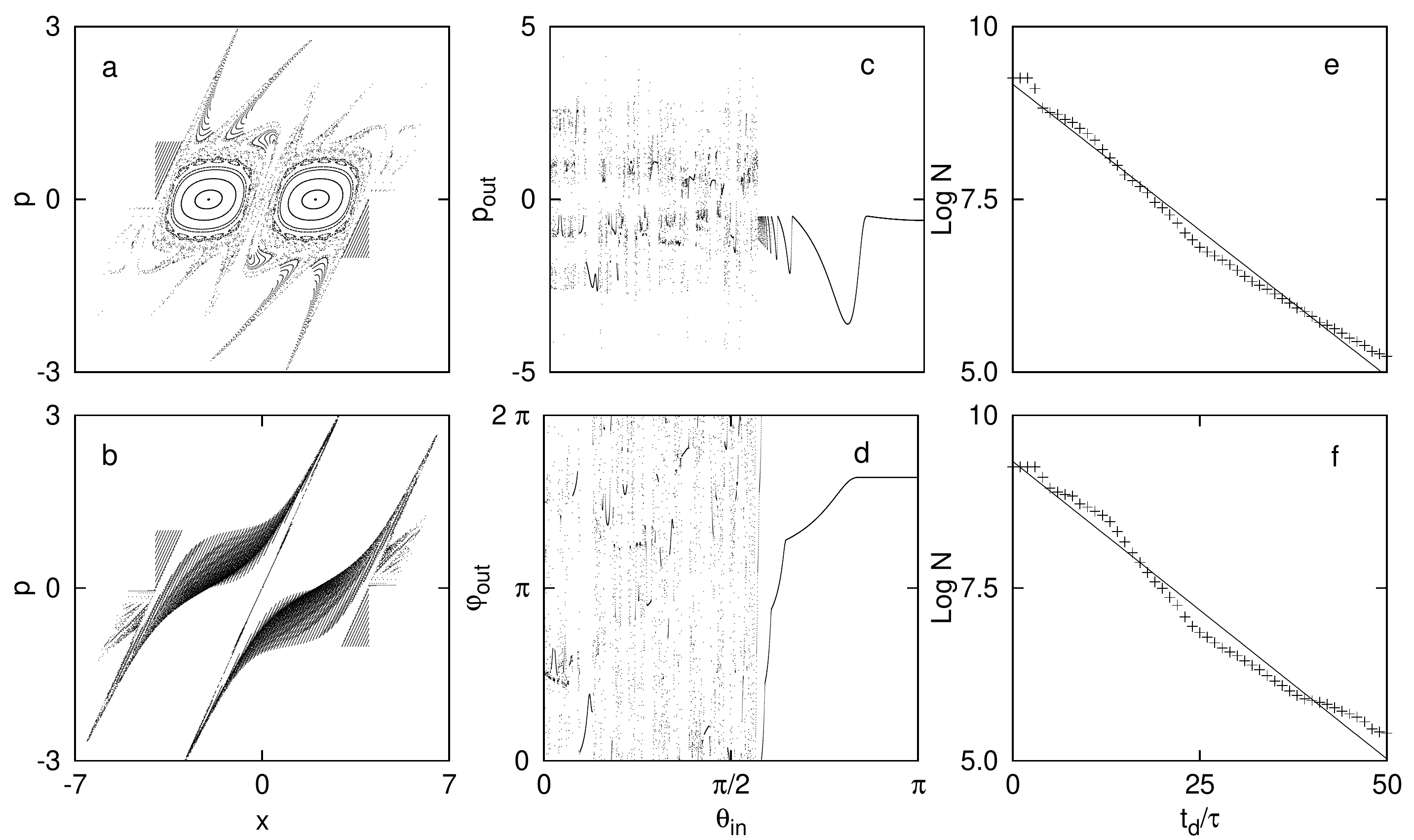}
\end{center}
\vspace{-0.5cm}
\caption{Irregular scattering in the classical angular momentum pump, Eqs.~(\ref{clhamiltonian}-\ref{bellies}). Left: Poincar\'e sections in the $(x,p)$-plane for positive ($\theta_{\rm
in}=0$, a) and negative incoming angular momentum ($\theta_{\rm in}=\pi$, b). Center:
Deflection functions for outgoing momentum $p_{\rm{out}}$  (c) and azimuth $\varphi_{\rm out}$(d) vs.\ initial polar angle $\theta_{\rm in}$. Right: Sojourn time statistics for
asymmetric ($A_1=2$, $A_2=1$, e) and symmetric field envelope ($A_1=A_2=1$, f). Straight lines: exponential decay for mean sojourn time $\langle t_{\rm stay}/T\rangle = 11.6$ (panel e) and 11.8 (f). Further initial conditions and parameters are $\varphi_{\rm in}=0$, $A_1=A_2=1$, and (c,d) $p_{\rm in}=1$, $\theta_{\rm in}=\pi/4$, $x_{\rm in}=-4$.}
\label{Fig:disp}
\end{figure}

As expected for a periodically driven system with two freedoms, Eqs.~(\ref{mapa}-\ref{mapc}) generate a non-integrable dynamics with a mixed phase space, see Fig.~\ref{Fig:disp}a,b. Criteria for irregular scattering \cite{Smi92}, such as fractal structures in the deflection functions (Figs.~\ref{Fig:disp}c,d) and an exponential distribution of sojourn times (Figs.~\ref{Fig:disp}e,f), are fulfilled. We observe in Figs.~\ref{Fig:disp}c,d that irregular scattering prevails for incoming angular momenta polarized in the direction of the field, $\theta_{\rm in} \lesssim \pi/2$, while for $\theta_{\rm in} \gtrsim \pi/2$, scattering is almost exclusively regular. This marked contrast is readily explained: For particles with $\cos(\theta_{\rm in}) < 0$ (spin down), the interaction $\gamma\,\textbf{s}\cdot\textbf{B}(x)$ amounts to a potential barrier reflecting them back before they enter the scattering region. By contrast, particles with $\cos(\theta_{\rm in}) > 0$ (spin up) see a potential well, are attracted into the scattering region, and undergo chaotic scattering which tends to randomize the outgoing with respect to the incoming conditions. The threshold appears slightly above $\theta_{\rm in} = \pi/2$, since for small negative $\textbf{B}\cdot\textbf{s}$ the incoming kinetic energy may still suffice to surmount the potential barrier.

This asymmetry is reflected in an imbalance between reflection and transmission processes, which in turn depends on the incoming direction and thus can be exploited to generate directed currents. If for individual scattering events the outgoing direction is the same, irrespective of the sign of $p_{\rm in}$, for otherwise equivalent incoming conditions, transport into that direction is preferred. Averaging over incoming directions and conditions, this may lead to biased probabilities for transport to the right vs.\ to the left. We calculate the probability current as \cite{DGS03}
\begin{equation}\label{I}
I_{\rm p} = \left(T_{\rm lr} + R_{\rm rr} - R_{\rm ll} - T_{\rm rl}\right),
\end{equation}
where $T_{\alpha\beta}$, $R_{\alpha\alpha}$, $\alpha,\beta = \textrm{l}$ (left) or $\textrm{r}$ (right), denotes the fraction of particles transmitted from channel $\alpha$ to $\beta$ or reflected from $\alpha$ back into $\alpha$, resp.

\begin{figure}[h!]
\begin{center}
\includegraphics[width=7cm]{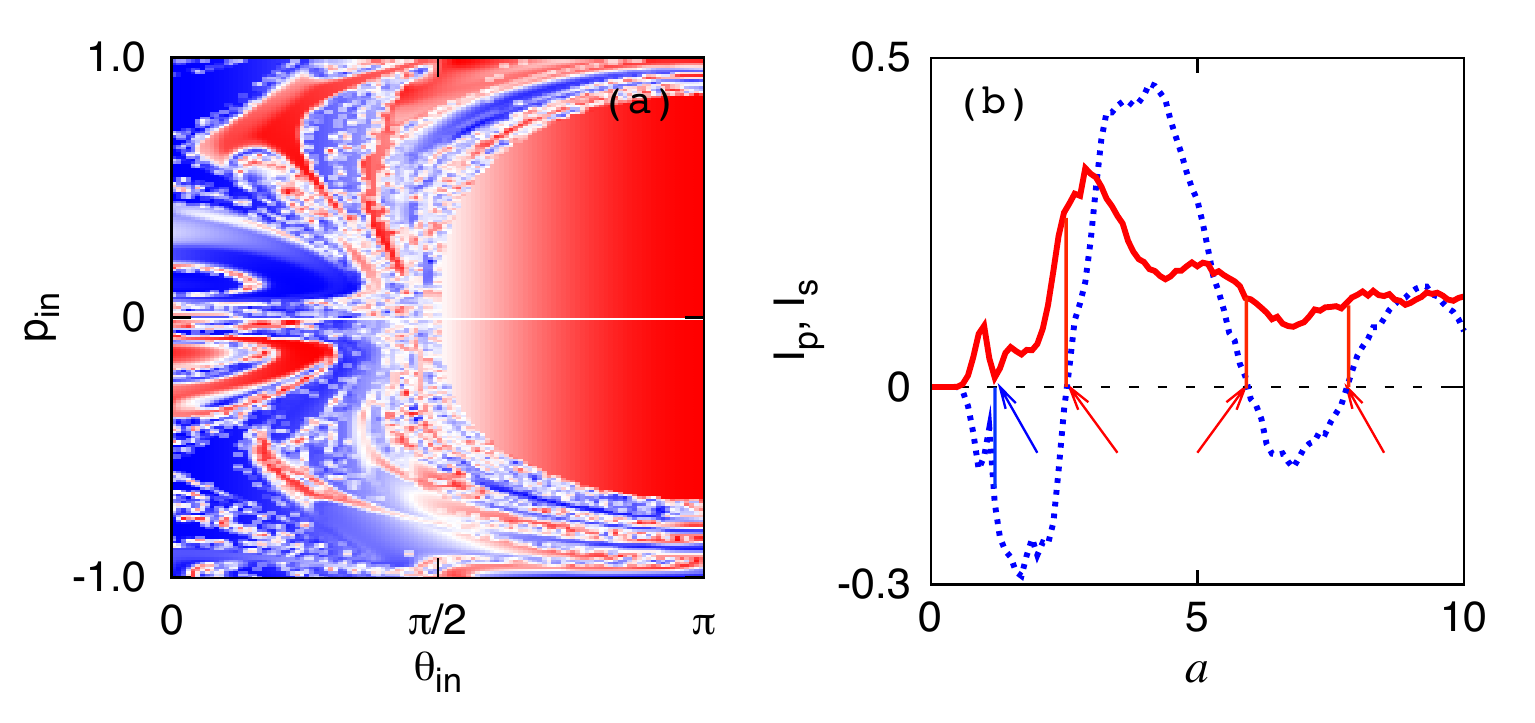}
\end{center}
\vspace{-0.7cm}
\caption{Directed currents in the classical angular momentum pump, Eqs.~(\ref{clhamiltonian}-\ref{bellies}). Panel a: Effective outgoing spin $\cos\theta_{\rm out}$ (color code, from red ($< 0$) through white ($0$) through blue ($> 0$)) as a function of incoming polar angle $\theta_{\rm in}$ and linear momentum $p_{\rm in}$, for 
$A_1 = 2$, $A_2 = 1$. Panel b: Particle current $I_{\rm p}$ (Eq.~(\ref{I}), dotted blue) and angular-momentum current $I_{s}$ (Eq.~(\ref{Isi}), red solid), as functions of the field width $a$ for  $A_1 = 8$, $A_2 = 1$, averaged over the ranges $0 \leq \theta_{\rm in} \leq \pi$ and $|p_{\rm in}| \leq 1$. Other initial conditions are $\phi_{\rm in} = \varphi_{\rm in} = 0$. Red (blue) arrows indicate pure spin (particle) transport.
For $A_1 = A_2 = 1$, both currents vanish (dashed).}
\label{Fig:corr}
\end{figure}

Angular-momentum currents will be defined, anticipating their generalization to spin currents, as $I_{\rm s} = I_+ - I_-$, where $I_+$ ($I_-$) are the partial currents for spin up (down). Replacing the discrete spin orientation by the continuous polar angle of the angular momentum, we arrive at the definition \cite{DD08},
\begin{equation}\label{Isi}
I_{s} = \frac{|\textbf{s}|}{4\pi} \int_0^\pi {\rm d}\theta \sin\theta
\int_{-\pi}^\pi {\rm d}\varphi \cos\theta \, j(\theta,\varphi),
\end{equation}
with the probability current density
$ j(\theta,\varphi) = j_{\rm lr}(\theta,\varphi) +  j_{\rm rr}(\theta,\varphi) - j_{\rm ll}(\theta,\varphi) - j_{\rm rl}(\theta,\varphi)$.
Partial currents $j_{\alpha\beta}$ are labelled in the same way as $T_{\alpha\beta}$, $R_{\alpha\alpha}$, cf.\ below Eq.~(\ref{I}), and are obtained by averaging over suitable ensembles of initial conditions with $(\theta,\varphi)$ fixed. Figure \ref{Fig:corr}a shows the effective outgoing spin per scattering event for broken reflection symmetry $A_1 \neq A_2$.



Particle as well as spin currents for the system (\ref{clhamiltonian}-\ref{bellies}) are shown in Fig.~\ref{Fig:corr}b as functions of the width $a$ of the scattering region, with and without symmetry  $A_1 = A_2$. Current reversals give rise to pure charge transport where the angular momentum current vanishes, and {\it vice versa\/} (blue and red arrows, resp.). 

\emph{Quantum spin pump.--} In the sequel we shall contrast classical chaotic angular-momentum pumps with the pumping of spin-$\frac{1}{2}$-particles, quantizing the spatial motion as well as the angular momentum of the model presented above. While jumping directly to the deep quantum regime as concerns the angular momentum, we leave the relative Planck's constant (the ratio of $\hbar$ to some characteristic action related to the $(x,p)$-phase space) as a parameter for the orbital motion. 

The Hilbert space appropriate for this setup comprises spinors $\boldsymbol{\psi}(x,t) = 
(\psi_-,\psi_+)$, $\psi_\pm(x,t)$ denoting the spin-up/down components of the wave function. Its time evolution is determined by the Schr\"odinger equation
${\rm{i}} \hbar \partial |{\boldsymbol{\psi}}(t)\rangle / \partial t = \hat{H}(t)|{\boldsymbol{\psi}}(t)\rangle$ with the Hamiltonian
\begin{equation}\label{qmhamiltonian}
\hat{H}(\hat{p},\hat{x},t) = \hat{H_0}(\hat{p}) + \hat{V}(\hat{x},t)
\sum_{n=-\infty}^{\infty}\delta(t-nT-t_{\rm{in}}),\\
\end{equation}
Here, $\hat{V}(\hat{x},t) = \mu_{\rm{B}}\hat{\boldsymbol{\sigma}} \cdot \textbf{B}(\hat{x})$, with $\hat{\boldsymbol{\sigma}} = (\hat\sigma_x,\hat\sigma_y,\hat\sigma_z)$, the vector of Pauli matrices. 
The Floquet operator
\begin{equation}
\hat U_{\rm{F}} =
\hat{\mathcal{T}} \exp\left(-\frac{\rm{i}}{\hbar}\int_{-0^+}^{T-0^+} {\rm{d}}t \,\hat{H}(t)\right),
\end{equation}
where $\hat{\mathcal{T}}$ effects time ordering, generates the time evolution over a single period $T$. With the kicked modulation (\ref{bellies}) of the magnetic field, it takes the form
\begin{equation}\label{qmap}
\hat U_{\rm{F}} = \exp\left(-\frac{\rm{i}}{\hbar} \hat{H_0} T\right)
\exp\left(\frac{\rm{i}}{\hbar}\mu_{\rm{B}} (B_1(\hat{x})\hat{\sigma}_y +
B_2(\hat{x})\hat{\sigma}_z)\right),
\end{equation}
which couples the spinors $\psi_-$ and $\psi_+$ to one another.
%
%

\begin{figure}[h!]
\begin{center}
\includegraphics[width=7cm,angle=0]{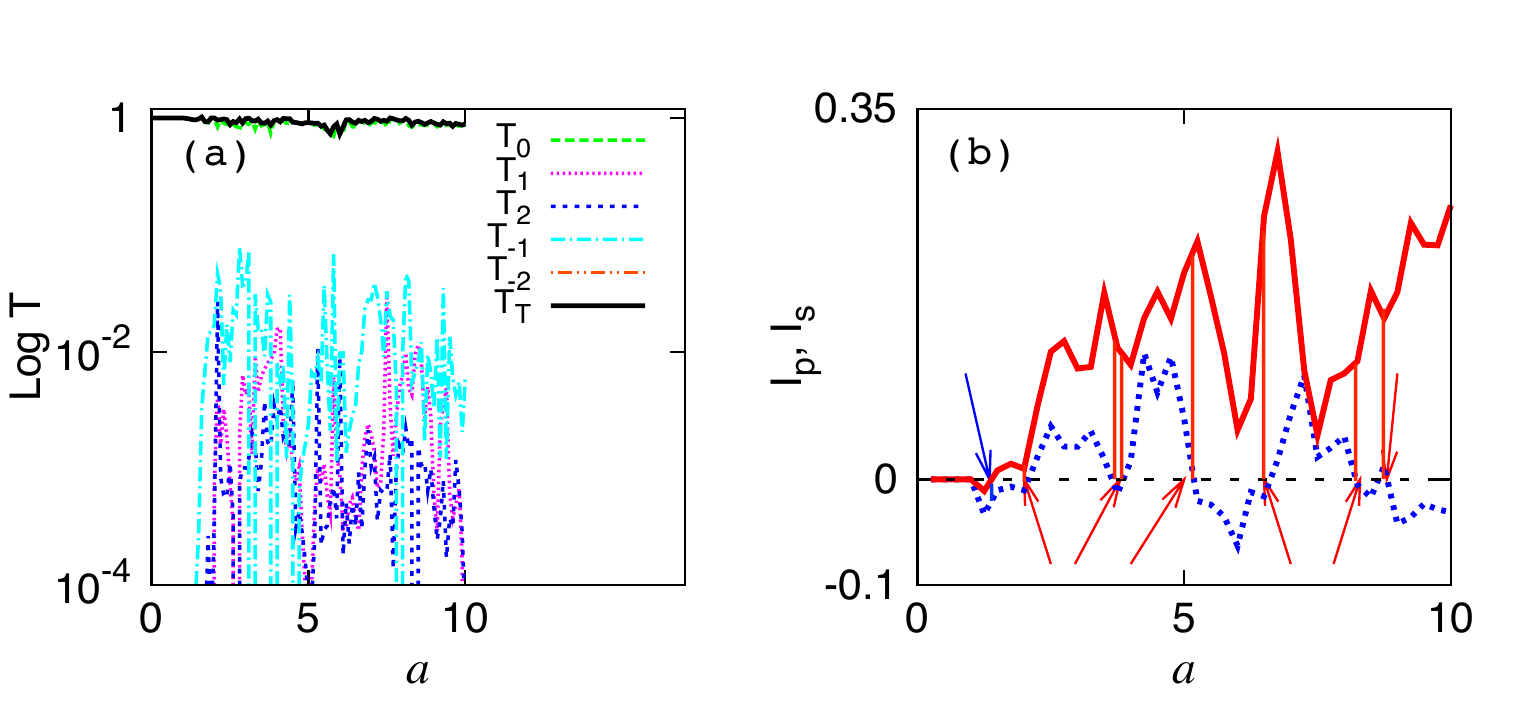}
\end{center}
\vspace{-0.7cm}
\caption{
Transmissions $T_{{\rm lr},l}$ per Floquet channel for $l = 0$, $\pm 1$, $\pm 2$ (Eq.~(\ref{TFlo}), panel a) and particle current (b, blue dotted curve) vs.\ spin current (b, red solid curve), as functions of the field width $a$ (cf.\ Eq.~(\ref{bellies})), for $\hbar = 0.25$, $A_1 =  8$, $A_2 =  1$, and $p_{\rm{in}} = 5$. Red (blue) arrows indicate pure spin (particle) transport.
For $A_1 = A_2 = 1$, both currents vanish (dashed).
}
\label{Fig:quantcu}
\end{figure}

Time-periodic scattering is inelastic in the sense that energy is conserved only ${\rm{mod}}\,\hbar\omega$, incoming and outgoing energies are related by
$E_{{\rm{out}},l}=E_{\rm{in}}+ l \hbar \omega$, with $\omega=2\pi/T$.
The index $l \in \mathbb{Z}$ labels Floquet channels and counts the number of photons exchanged with the driving field \cite{HDR01}. Likewise, transmission and reflection coefficients can be decomposed in Floquet channels, e.g., by Fourier transforming the asymptotic outgoing waves $\psi_{\pm\infty}(p)$.
The transmission probability per channel is then obtained as
\begin{equation}\label{TFlo}
T_{{\alpha\beta},l}(E_{\rm{in}})=\left|\frac{p_{\rm{in}}}{p_{{\rm{out}},l}}\frac{\psi_{\infty}(p_{{\rm{out}},l})}{\psi_{-\infty}(p_{\rm{in}})}\right|^{2},
\end{equation}
where $p_{{\rm{in}}({\rm{out}},l)} = \sqrt{2mE_{{\rm{in}}({\rm{out}},l)}}$. Figure \ref{Fig:quantcu}a shows the transmissions $T_{{\rm lr},l}$ for $l = 0$, $\pm 1$, $\pm 2$ vs.\  the width $a$ of the field sectors. They have been obtained by scattering a continuous incoming wave \cite{JC&08,Dub10}, which by contrast to wavepacket scattering allows to precisely define the incoming momentum $p_{\rm{in}}$ and hence to break down the outgoing plane waves unambiguously into Floquet channels $\psi_{\pm\infty}(p_{{\rm{out}},l})$ \cite{Dub10}.

In figure \ref{Fig:quantcu}b, we compare particle and spin currents as functions of the width $a$,  for fixed field amplitudes $A_1$, $A_2$.
As in the classical case, reversals of the particle current $I_{p}$ at appreciable values of the spin current $I_{s}$ (red arrows) give rise to pure spin transport. The r\^ole of symmetry breaking is clearly reflected in the transport features: In the symmetric case $A_1 = A_2$ (dashed), both currents vanish identically. They tend to vanish as well for too low driving frequency, corresponding to the adiabatic limit where a two-parameter driving would be required to generate directed transport. Similarly, for too strong driving, the mechanism for asymmetric scattering pointed out above looses validity and pure spin currents do not appear any more.

\emph{Conclusion.--} As a prototypical example for periodically driven chaotic scattering with internal freedoms, we have presented a model that couples the angular momentum of the scattered particles via a time-dependent inhomogeneous magnetic field to the orbital motion. The magnetic field serves three purposes, breaking time-reversal invariance, rendering the motion chaotic, and inducing spin flips. Particle as well as angular momentum currents generated by the irregular scattering exhibit frequent current reversals which give rise to pure transport of charge or angular momentum. The markedly asymmetric transport is explained by the fact that only for one spin polarization, particles undergo irregular scattering, while for the opposite orientation, they are immediately reflected. This mechanism does not require a driving that breaks TRI, a symmetric single-parameter time-dependence is sufficient.

These results largely carry over to spin pumps, obtained by quantizing angular momentum pumps. Representing spin-${1\over 2}$-particles by spinors and treating the periodic driving in the Floquet formalism, we derive a quantum map that couples spin flips to the orbital motion. As in classical asymmetric scattering, we find directed transport with current reversals both for charge and spin currents. It is even possible to decompose these currents in Floquet channels, associating directed transport to the exchange of photons with the driving field. Non-vanishing spin transport at zeros of the charge current opens a new way, based on nonlinear dynamics, to generate polarized currents, applicable, e.g., in spintronics. Implementing the pump as a semiconductor superlattice (spatial scale $\sim\! 10\,$nm), with the kicked electromagnetic field coupled in as a modulated (e.g., as a frequency comb) THz free-electron laser source \cite{GK&93} (kick period $\sim\! 10^{-11}\,$s, peak magnetic field $\sim\! 0.16\,$T), would result in interaction energies of the order of $10^{-6}$eV for electrons. It is comparable to the asymptotic kinetic energy of incoming electrons, for the same parameters and spatial and temporal scales of our model, indicating that an experimental realization with state-of-the-art laboratory equipment is feasible.

The technically demanding task of numerical simulations in the semiclassical regime of the orbital degree of freedom, as well as corresponding semiclassical approximations \cite{CS&04,Cul09}, remain as challenges for future work.

%
We gratefully acknowledge financial support by Volks\-wagen Foundation (grant I/78235) and Colciencias (grant 1101-05-17608). We enjoyed fruitful discussions with F.~Leyvraz and K.~Richter and appreciate the hospitality of the Institute for Theoretical Physics during research stays of TD and FLD at University of Regensburg.

\section*{References}

\bibliographystyle{unsrt}

\end{document}